\documentclass{PoS}

\title{Meson masses and decay constants at large $N$}

\ShortTitle{Meson masses and decay constants at large $N$}

\author{ Margarita Garc\'{\i}a P\'erez$^{a}$, Antonio Gonz\'alez-Arroyo
$^{a,b}$ and \speaker{Masanori Okawa}$^{c,d}$ \\
  $^a$ Instituto de F\'{\i}sica Te\'orica UAM-CSIC, Nicol\'as Cabrera 13-15, \\
  Universidad Aut\'onoma de Madrid, E-28049--Madrid, Spain \\
  $^b$ Departamento de F\'{\i}sica Te\'orica, C-15 \\
       Universidad Aut\'onoma de Madrid, E-28049--Madrid, Spain \\
  $^c$ Graduate School of Science, Hiroshima University,\\
Higashi-Hiroshima, Hiroshima 739-8526, Japan \\
  $^d$ Core of Research for the Energetic Universe, Hiroshima University,\\
Higashi-Hiroshima, Hiroshima 739-8526, Japan \\

E-mail: \email{margarita.garcia@uam.es, antonio.gonzalez-arroyo@uam.es, okawa@sci.hiroshima-u.ac.jp}
 }

\abstract{Meson masses and decay constants in the large $N$ limit of SU($N$) gauge theory are determined using the twisted Eguchi-Kawai
reduced model. 
To this end, we make use of a recently defined smearing method valid on the one-point lattice. 
This procedure, in combination with a variational analysis,  allows to obtain reliable values for these quantities.
} 

\FullConference{34th annual International Symposium on Lattice Field Theory\\
		24-30 July 2016\\
		University of Southampton, UK}

\newcommand{\be}{\begin{equation}}
\newcommand{\ee}{\end{equation}}
\newcommand{\ba}{\begin{array}}
\newcommand{\ea}{\end{array}}
\newcommand{\baa}{\begin{array}}
\newcommand{\eaa}{\end{array}}
\newcommand{\bea}{\begin{eqnarray}}
\newcommand{\eea}{\end{eqnarray}}

\begin{document}

\section{Introduction}

The Twisted Eguchi-Kawai model (TEK-model) is a $SU(N)$ lattice gauge theory having only one site with twisted boundary conditions~\cite{TEK1,TEK2}. 
It has been shown that this model is equivalent to the usual lattice gauge theory in the large $N$ limit~\cite{TEK3,testing}.  
In fact, we have succeeded in calculating the large $N$ string tension and running coupling constant in the continuum limit starting  
from the TEK model~\cite{string,coupling}.  We have also shown that it is straightforward to introduce adjoint fermions 
within our framework~\cite{nf2prd,nf2jhep}. Last year, two of us proposed a new method to calculate meson correlators in this set up
~\cite{meson,mesonlat15}. This problem was actually quite challenging for two main reasons:   
\begin{itemize}
\item Quarks in the fundamental representation are not consistent with the twist.
\item Meson propagators are space-time extended objects not easily defined within the one-site lattice model.  
\end{itemize}  
Both caveats can be solved by allowing quarks to propagate in a bigger lattice than the one seen by the periodic (up to the twist) 
gauge fields. This has allowed to compute the large $N$ ground state meson spectrum, using for simplicity point-point meson 
correlators~\cite{meson,mesonlat15}.  However, it was not possible to reliably estimate decay constants due to the contamination from excited 
states. The purpose of the present talk is to make use of the smearing method introduced in~\cite{mesonlat15}, which combined with a variational analysis makes it possible to 
calculate both meson masses and decay constants.

\section{Formulation}

In this section we will briefly review the construction of the meson correlators in the TEK model, further details can be consulted in
refs.~\cite{meson,mesonlat15}. We formulate the $SU(N)$ gauge theory, with $N=L^2$, on a 4-dimensional one-point lattice with twisted boundary conditions given 
by the antisymmetric twist tensor: $n_{\nu\mu} = k L$, for $\mu>\nu$. The integers $L$ and $k$ should be coprime and obey certain constraints 
to prevent center symmetry breaking~\cite{TEK3}.   
In the large $N$ limit, this theory is equivalent to the usual $SU(N)$ lattice gauge theory defined on a $L^4$ lattice. 

We will consider that quark fields live on a finite box $\ell_0 L \times L^3$, with positive integer $\ell_0$. In ref.~\cite{meson}, we have shown
that the correlation function at time separation $n_0$ of two local, zero-momentum meson operators in channels $\gamma_A$ and $\gamma_B$  is 
given by:
\begin{equation}
C_{AB}(n_0)={1 \over \ell_0 N^{3/2}} \sum_{q_0}{\rm e}^{-i q_0 n_0} {\rm Tr} \left[\gamma_A D^{-1}(0)\gamma_B D^{-1}(q_0)\right],
\label{meson_prop}
\end{equation}
with momentum $q_0$ quantized in units of $\ell_0 L$. The Wilson-Dirac operator acts on colour ($U_\mu$), spatial ($\Gamma_\mu$), and Dirac ($\gamma_\mu$)
indexes, and is given by: 
\begin{equation}
 D(q_0)=1-\kappa\sum_{\mu=0}^{d-1}\left[(1-\gamma_\mu){\tilde U_\mu} \Gamma_\mu^{*} +
(1+\gamma_\mu){\tilde U_\mu^\dagger} \Gamma_\mu^{t} \right],
\label{quark_action}
\end{equation}
with $ {\tilde U_{\mu}}=\exp(i q_\mu \delta_{\mu, 0})\, U_{\mu}$,
and with $\Gamma_\mu$ the $SU(N)$ matrices satisfying 
\be
 \Gamma_\mu \Gamma_\nu = \exp(2 \pi i n_{\nu\mu}/N)   \Gamma_\nu \Gamma_\mu.
\ee
An explicit form for these matrices, usually called twist eaters, is shown, for example, in eq.(2.13) of ref.~\cite{testing}. 

Although one can compute meson masses from these local correlators, the effective
mass plots show a clear contamination from excited states~\cite{mesonlat15}.  To obtain improved results, we have proposed to use the smearing method originally introduced in~\cite{Bali} adapted to the one-point lattice~\cite{mesonlat15}.  
Smearing can be easily implemented by replacing $\gamma_A$ in eq.~(\ref{meson_prop}) by the operator:
\begin{equation}
\gamma_A \rightarrow D_s^l \gamma_A,\ \ \ D_s \equiv {1 \over 1+6c}\left[1+c \sum_{i=1}^{d-1}\left( {\bar U_i} \Gamma_i^{*} +
{\bar U_i^\dagger} \Gamma_i^{t} \right) \right].
\label{smearing}
\end{equation}
Here, $l$ is the smearing level and ${\bar U_i}$ is the APE-smeared spatial link variable obtained 
after iterating several times the following transformation
\begin{equation}
U'_i = {\rm Proj_{SU(N)}} \left[ (1-f)U_i + {f \over 4} \sum_{j \ne i} ( U_j U_i U_j^{\dagger} + U_j^{\dagger} U_i U_j) \right]\ ,
\label{ape}
\end{equation}
with $c$ and $f$ free smearing parameters. 

\begin{figure}[t]
 \begin{minipage}{0.5\hsize}
  \begin{center}
   \includegraphics[width=65mm]{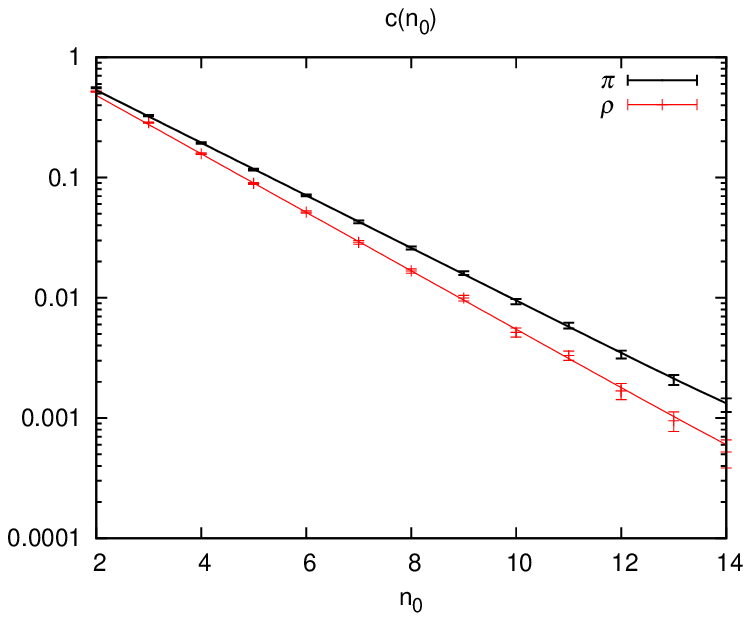}
  \end{center}
\vspace{-0.7cm} 
  \caption{$SS$ correlators for $\pi$  and $\rho$.}
  \label{fig_SS_corr}
 \end{minipage}
 \begin{minipage}{0.5\hsize}
  \begin{center}
   \includegraphics[width=65mm]{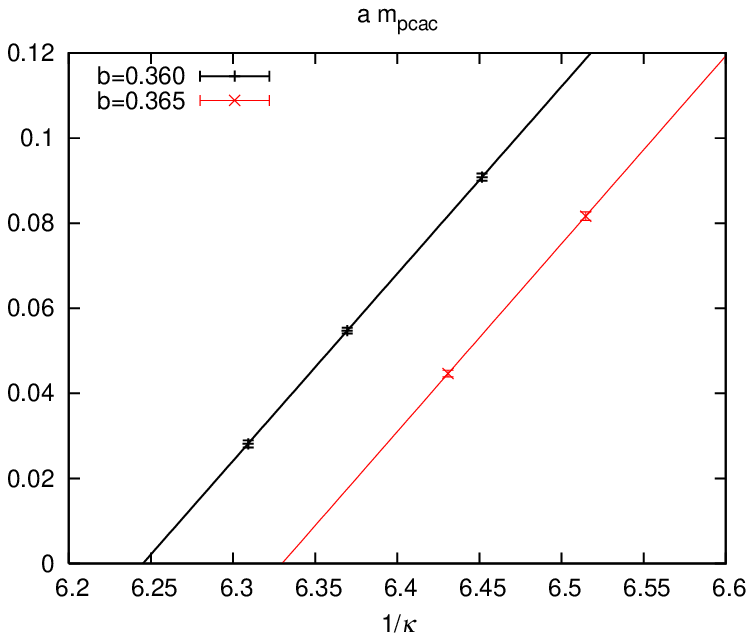}
  \end{center}
\vspace{-0.7cm}
  \caption{$1/\kappa$ dependence of $am_{\rm PCAC}$.}
  \label{fig_k_vs_m_pcac}
 \end{minipage}
\end{figure}

To improve the ground state signal we have used in addition a variational method. 
We start by selecting a basis of operators in each quantum channel. This is done for fixed
APE-smearing of the gauge fields by changing the smearing level $l$ of the fermionic operator.
We have selected ten smearing steps $\ell=$ 0, 1, 2, 3, 4, 5, 10, 20, 50, 100, and computed the $10\times 10$ correlation matrices
$C_{AB}(n_0)$ with matrix components: 
\begin{equation}
C_{AB}^{ll'}(n_0)={1 \over \ell_0 N^{3/2}}\sum_{q_0}{\rm e}^{-i q_0 n_0} {\rm Tr} \left[D_s^l \gamma_A D^{-1}(0)D_s^{l'} \gamma_B D^{-1}(q_0)\right].
\label{smeared_meson_prop}
\end{equation} 
We then solve the generalized eigenvalue problem (no summation over $A$ implied):
\begin{equation}
C_{AA}(n_0=t_2)v_{A}=\lambda_{A}C_{AA}(n_0=t_1)v_{A},
\label{generalized_eigen}
\end{equation}
with $t_1=1$ and $t_2=2$. After this, we select the eigenvector
corresponding to the largest eigenvalue in each channel, denoted   
by $v_A^1$.  Smeared-smeared and local-smeared correlators are then constructed for arbitrary $n_0$ as  
\begin{eqnarray}
C_{AB}^{SS}(n_0)&=& \sum_{\ell, \ell'}\ \  (v_A^1)^*_\ell \ \ C_{AB}^{\ell \ell'}(n_0) \  (v_B^1)_{\ell'}, \\
C_{AB}^{LS}(n_0)&=&   \sum_{\ell'} \ \ C_{AB}^{0 \ell'}(n_0) \  (v_B^1)_{\ell'}.
\end{eqnarray}
In the following, $SS$ correlators will be used to determine meson masses, while $LS$ correlators will allow for the determination
of decay constants in a way to be described below.

We show in fig.~\ref{fig_SS_corr} the $SS$ pion and rho correlators for $N=289$ and $l_0=2$, at $b=0.36$ and $\kappa=0.155$,
from which a precise determination of the masses can be obtained. 
For this plot, the smearing parameters have been set to $c = 0.4$, $f = 0.15$ and we
performed 10 Ape-iterations of the gauge fields. 
We are currently working on a systematic analysis of the effects associated to the choice of operators in the
correlation matrix. In the range of $\kappa$ and $\beta$ values of our simulations a basis of 8 to 10 smearing
operators seems to be optimal for reducing the excited state contamination without loosing precission in the effective masses.
A next step involves solving the generalized eigenvalue problem at all values of $t_2$ in eq.~(\ref{generalized_eigen}), 
as discussed in~\cite{sommer}.

\section{Simulations and results}

To test our method, we have simulated the TEK model for  $N$=289, with  $\ell_0=2$.
We took two values of inverse 't Hooft coupling ($b=1/g_0^2N$) and two or three values of $\kappa$,
\begin{eqnarray}
b&=&0.360,\ \ \ \kappa=0.1555, 0.157, 0.1585, \\
b&=&0.365,\ \ \ \kappa=0.1535, 0.1555.
\end{eqnarray}
For each parameter set, we have calculated meson propagators with 800 configurations, each configuration being separated by 1000 MC sweeps~\cite{OR}.
In the following, all meson masses are obtained from a single hyperbolic cosine fit to the $SS$ correlator in each channel, with the fitting range 
$6 \le n_0 \le 12$.

The first quantity we study is the PCAC-quark mass.  Following ref.~\cite{Bali}, we compute it from a constant fit to:
\begin{equation}
am_{\rm PCAC}(n_0) = { C^{LS}_{\gamma_0 \gamma_5,\gamma_5}(n_0+1) -C^{LS}_{\gamma_0 \gamma_5,\gamma_5}(n_0-1) \over 4 C^{LS}_{\gamma_5,\gamma_5}(n_0) }.
\label{pcac}
\end{equation}

\begin{figure}[t]
 \begin{minipage}{0.5\hsize}
  \begin{center}
   \includegraphics[width=65mm]{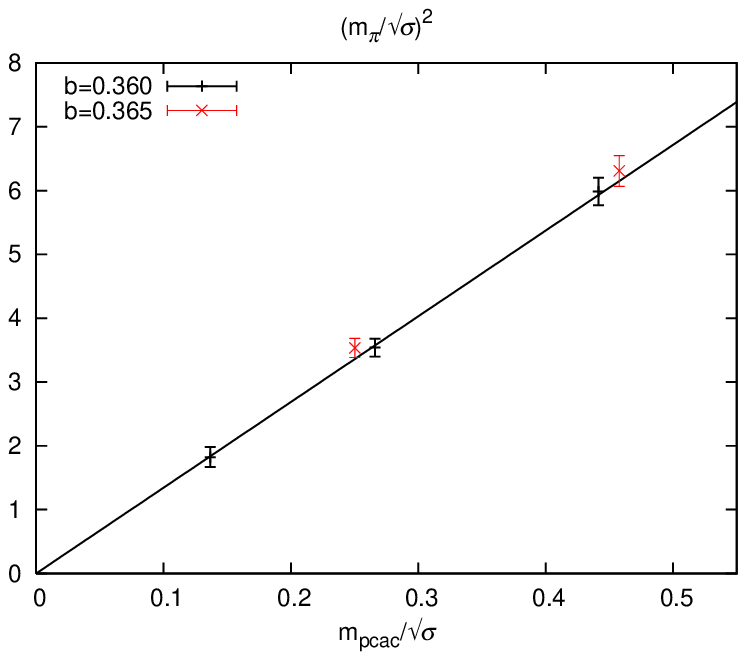}
  \end{center}
\vspace{-0.7cm}
  \caption{$m_{\rm PCAC}/\sqrt{\sigma}$ dependence of $(m_\pi/\sqrt{\sigma})^2$.}
  \label{fig_m_pcac_m_pi^2}
 \end{minipage}
 \begin{minipage}{0.5\hsize}
  \begin{center}
   \includegraphics[width=65mm]{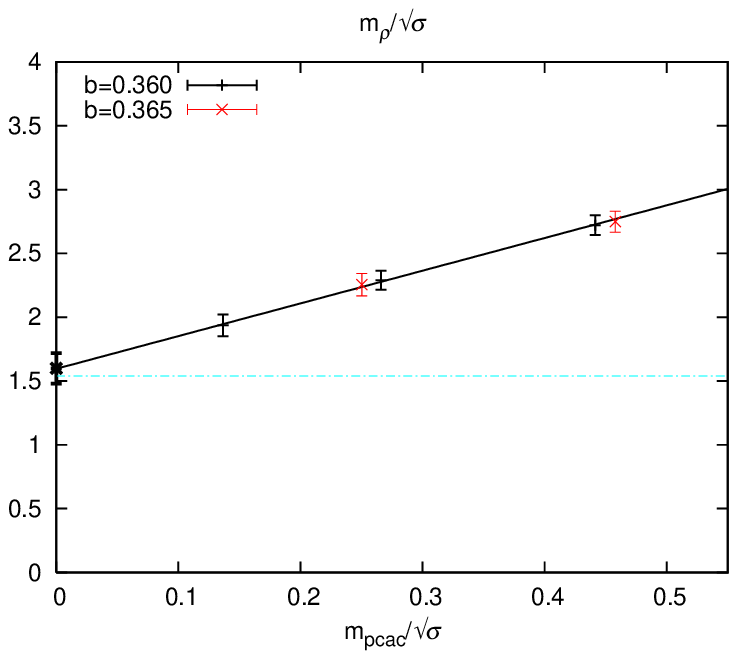}
  \end{center}
\vspace{-0.7cm}
  \caption{$m_{\rm PCAC}/\sqrt{\sigma}$ dependence of $m_\rho/\sqrt{\sigma}$.}
  \label{fig_m_pcac_m_ro}
 \end{minipage}
\end{figure}

In fig.~\ref{fig_k_vs_m_pcac}, we show the $1/\kappa$ dependence of $am_{\rm PCAC}$. In the case were we have three values of $\kappa$ a linear fit provides a very good
fit to the data. Fig.~\ref{fig_m_pcac_m_pi^2} displays  $(m_\pi/\sqrt{\sigma})^2$  as a function of $m_{\rm PCAC}/\sqrt{\sigma}$. The value of string tension 
is extracted from ref.~\cite{string}, and is given by $a^2\sigma= 0.04234(103)$ and $a^2\sigma= 0.03181(60)$
for $b=0.36$ and $b=0.365$ respectively. The straight line is a linear fit to the data at $b$=0.360 setting the pion residual mass to zero. 
It is important to note that in the large $N$ limit, there should be 
no quenched chiral log~\cite{BG,sharpe}.  Therefore the verification of the linear dependence of $(m_\pi/\sqrt{\sigma})^2$ 
on $m_{\rm PCAC}/\sqrt{\sigma}$ is a non-trivial consistency check that we are really studying the dynamics of large $N$ QCD. 

In fig.~\ref{fig_m_pcac_m_ro}, we show the dependence of $m_\rho/\sqrt{\sigma}$ on $m_{\rm PCAC}/\sqrt{\sigma}$.  Again, the straight line is a linear fit of 
the data at $b$=0.360. The results at $b$ =0.36 and 0.365  exhibit a very good scaling behavior. 
The horizontal line in fig.~\ref{fig_m_pcac_m_ro} is the value of $m_\rho/\sqrt{\sigma}$ given in~\cite{Bali} which is obtained by extrapolating $m_\rho$ at 
finite $N$ to $N \to \infty$, and comes out very consistent with our determination. 
Scaling violations seem to be really small, taking into account that no continuum extrapolation has been performed in either case.

\begin{figure}[t]
 \begin{minipage}{0.5\hsize}
  \begin{center}
   \includegraphics[width=65mm]{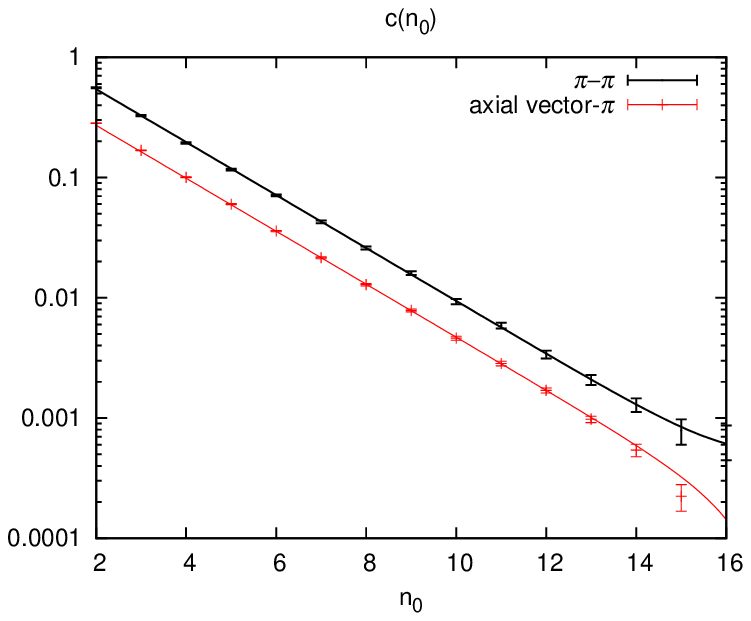}
  \end{center}
\vspace{-0.7cm}
  \caption{$\gamma_5$-$\gamma_5$, and $\gamma_0 \gamma_5$-$\gamma_5$ correlators.}
  \label{fig_axial}
 \end{minipage}
 \begin{minipage}{0.5\hsize}
  \begin{center}
   \includegraphics[width=65mm]{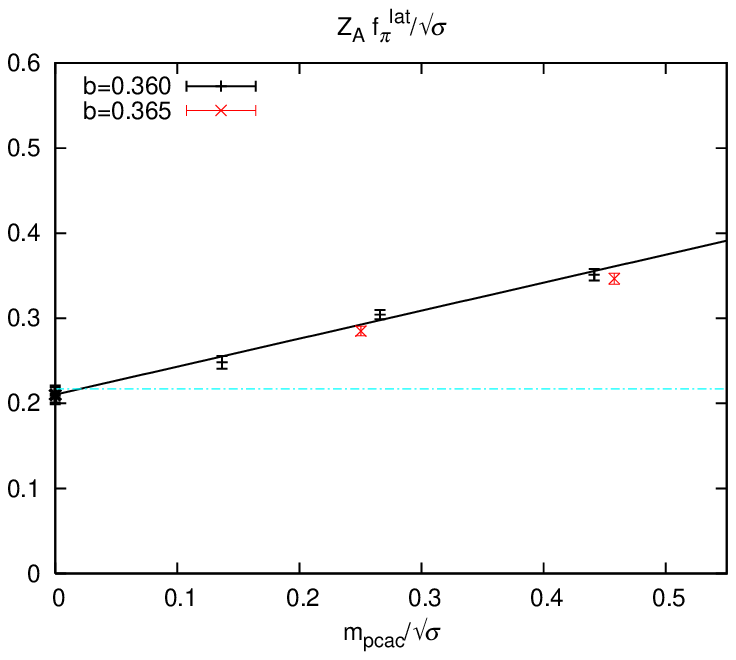}
  \end{center}
\vspace{-0.7cm} 
  \caption{$m_{\rm PCAC}/\sqrt{\sigma}$ dependence of $Z_A f_\pi^{\rm lat}/\sqrt{\sigma}$.}
  \label{fig_m_pcac_f_pi}
 \end{minipage}
\end{figure}

Finally, we calculate the pion decay constant $f_\pi^{\rm lat}$. For that we first compute the correlator between
the local axial-vector operator and the smeared pion operator: 
\be
C^{LS}_{\gamma_0 \gamma_5,\gamma_5}(n_0) =  C_{A_0} {\rm e}^{-n_0 m_\pi}.
\ee
This decays exponentially in time with the pion mass. The coefficient $C_{A_0}$ is related to the decay constant through:
\be
C_{A_0} = f_\pi^{\rm lat} \sqrt{{N \over 3} m_\pi C_\pi}
\ee
where $C_\pi$ can be extracted from the $SS$ pion correlator:
\be 
C^{SS}_{\gamma_5,\gamma_5}(n_0) = C_\pi {\rm e}^{-n_0 m_\pi}. 
\ee
We fit simultaneously these two correlators to hyperbolic sine and cosine functions respectively in the fitting range $6 \le n_0 \le 12$, keeping as free
parameters: $m_\pi$, $C_\pi$ and $f_\pi$. An example of the quality of our fits is shown on fig.~\ref{fig_axial},
for the same set of configurations as in fig.~\ref{fig_SS_corr}. The results for the pion mass come out compatible with the previous determination based only on the
$SS$ correlator.

This gives the unrenormalized pion decay constant. To connect with the continuum decay constant, we follow~\cite{Bali} and use the  
one-loop improved $Z_A$ renormalization factor~\cite{haris}
\begin{eqnarray}
&&Z_A= 1-0.4694{\lambda_E \over 4\pi},\ \ \ \lambda_E=-8{\rm ln} U_P, \\
&&Z_A(b=0.36)=0.8256,\ \ \ Z_A(b=0.365)=0.8314.
\end{eqnarray} 
The final result for the pion decay constant as a function of the PCAC quark mass is displayed in fig.~\ref{fig_m_pcac_f_pi}.  Both quantities are given
in units of the square root of the string tension. The results show a very good scaling behaviour and nicely extrapolate in the chiral limit
to the result obtained by standard large $N$ techniques in~\cite{Bali}, indicated by the blue horizontal line in the plot.

\section{Conclusions}
We have shown that our smearing method, when combined with a variational analysis, works quite well and 
allows to determine the  pion and rho masses, as well as the pion decay constant for the TEK model.
Our results are consistent with those obtained using standard  large $N$ techniques as in~\cite{Bali}.
   
In the future we plan to extend our results making  full use of the
generalized eigenvalue method to give a precise estimate of the masses and
decay constants in the continuum limit. These results will  include the evaluation  
of other channels such as scalar, tensor and
axial-vector. Other interesting extensions are possible and included
in the future plans. The study of large  $N$ 2-dimensional QCD seems
an interesting testing ground, since the meson mass spectra 
is known in that case as the solution  of an integral equation in the 
continuum limit~\cite{2dqcd}. Preliminary  results are presented  in these proceedings~\cite{GAO}. 
An interesting case is that of  $SU(N)$ field theories with  adjoint fermions
with varying  number of flavors, which has received much attention in connection with 
possible extensions of the Standard Model. The  configurations needed
for the determination of the meson spectrum are already available as a
by-product of our computation of  the mass anomalous dimension
for these theories~\cite{nf2prd,nf2jhep}.

\acknowledgments{
We acknowledge financial support from the 
grants FPA2012-31686, FPA2012-31880,  FPA 
2015-68541-P,
and the Spanish MINECO's ``Centro
de Excelencia Severo Ochoa'' Programme under grant
SEV-2012-0249. M. O. is supported by the Japanese MEXT grant No
26400249 and the MEXT program for promoting the enhancement of research
universities.
Calculations have been done on both Hitachi SR16000 supercomputer
at High Energy Accelerator Research Organization(KEK) and NEC SX-ACE supercomputer at Cybermedia Center, Osaka University
under the support of Research Center for Nuclear Physics(RCNP), Osaka University. We also acknowledge the use of the 
HPC resources at IFT.  Work at KEK is supported by the Large Scale Simulation Program No.16/17-02. 
}

\end{document}